\documentclass[prb,preprint,showpacs,eqsecnum]{revtex4-1}
\usepackage{graphicx}
\usepackage{bm}
\usepackage{amsmath,amssymb}
\begin{document}
\title{Energy dissipation in single-domain ferromagnetic nanoparticles: Dynamical approach}
\author{T.~V.~Lyutyy$^{1}$}
\email{lyutyy@oeph.sumdu.edu.ua}
\author{S.~I.~Denisov$^{1}$}
\email{denisov@sumdu.edu.ua}
\author{A.~Yu.~Peletskyi$^{1,2}$}
\author{C.~Binns$^{3}$}
\affiliation{$^{1}$Sumy State University, 2 Rimsky-Korsakov Street, UA-40007 Sumy, Ukraine\\
$^{2}$Institut f\"{u}r Physik, Universit\"{a}t Augsburg, Universit\"{a}tsstra{\ss}e 1, D-86135 Augsburg, Germany\\
$^{3}$University of Leicester, Leicester, LE1, 7RH, UK}

%\date{submitted to Physical Review B: \today}

\begin{abstract}
We study, both analytically and numerically, the phenomenon of energy dissipation in single-domain ferromagnetic nanoparticles driven by an alternating magnetic field. Our interest is focused on the power loss resulting from the Landau-Lifshitz-Gilbert equation, which describes the precessional motion of the nanoparticle magnetic moment. We determine the power loss as a function of the field amplitude and frequency and analyze its dependence on different regimes of forced precession induced by circularly and linearly polarized magnetic fields. The conditions to maximize the nanoparticle heating are also analyzed.
\end{abstract}
\pacs{75.50.Tt, 76.20.+q, 84.60.-h}
\maketitle

\section{INTRODUCTION}
\label{Int}

Single-domain ferromagnetic nanoparticles are of great interest due to their unique physical properties such as superparamagnetism,\cite{Neel, Brown} macroscopic quantum tunneling of magnetization,\cite{ThLi, ChTe} size-dependent characteristics,\cite{Fior} and exchange bias.\cite{NoSo, IgLa} These and other nanoscale properties of ferromagnetic nanoparticles make them very attractive for applications, e.g., in high density data storage,\cite{Ross, MTMA, Kiki} spintronic devices,\cite{WABD, ZFS, Shin} and biomedical engineering.\cite{PCJD, Ferr, LaLe, LFPR} If the magnetic state of nanoparticles is controlled by a time-dependent external magnetic field, then nanoparticles absorb energy from the field and heat up. While the heating of nanoparticles is undesirable for most applications, this property is of crucial importance for magnetic hyperthermia applications\cite{PCJD, LaLe, LFPR} (see also Refs.~\onlinecite{LiBi, MeTa, CoEs} and references therein). In a ferrofluid subjected to an external periodic magnetic field two heating mechanisms are usually considered,\cite{Shli, Rosen} one of which is related to Brownian rotation of nanoparticles and the other to their N\'{e}el relaxation. With these thermal-induced mechanisms, the energy dissipation per period is expressed in terms of the imaginary part of the magnetic susceptibility of the ferrofluid. It should be noted that since the magnetic susceptibility is a function of only the Brownian and N\'{e}el relaxation times,\cite{Rosen} the deterministic dynamics of the magnetic moment of nanoparticles does not affect the energy dissipation.

In contrast, if the rotation of nanoparticles is not allowed and the superparamagnetic state is not realized (this occurs, e.g., when the nanoparticles are embedded in a solid matrix and the temperature is small enough), then the heating phenomenon is expected to be strongly dependent on the dynamics of the nanoparticle magnetic moment. On a phenomenological level, it can be described by the deterministic Landau-Lifshitz-Gilbert (LLG) equation.\cite{LaLi, Gilb} Because of its nonlinearity, the precessional motion of the magnetic moment can be very complex. In particular, the circularly polarized magnetic field, whose polarization plane is perpendicular to the anisotropy axis, can generate the periodic and quasiperiodic regimes of precession of the magnetic moment.\cite{BSM, DLHT, SuWa, BMS} Moreover, the precessional motion of the magnetic moment induced by the linearly polarized magnet field can exhibit chaotic behavior.\cite{APC, VaPo, BPSV}

The aim of this paper is to study the dependence of energy dissipation on these regimes of precession of the nanoparticle magnetic moment. To the best of our knowledge, this problem has not been addressed before. We emphasize that the above-mentioned regimes and transitions between them can exist only in anisotropic nanoparticles. Note that the energy dissipation in such nanoparticles, arising from the precessional motion of the magnetic moment, was the subject of Ref.~\onlinecite{NaRa}. But the authors considered only the periodic regime of precession induced by the circularly polarized magnetic field. It is also important to stress that, because the influence of eddy currents on the magnetic moment dynamics can be accounted for by introducing an additional damping parameter,\cite{DLPB} the analysis presented below is applicable to both dielectric and metallic nanoparticles.

The paper is organized as follows. In Sec.~\ref{Desc}, we describe the model and introduce the reduced power loss. The analytical solutions of the LLG equation, obtained in the small amplitude approximation for both circularly and linearly polarized magnetic fields, and the corresponding power losses are presented in Sec.~\ref{An_PL}. Our numerical results are reported in Sec.~\ref{An_PL}. Here, the reduced power loss and its connection with the character of the precessional motion of the nanoparticle magnetic moment are studied depending on the amplitude and frequency of these magnetic fields. We summarize our findings in Sec.~\ref{Summ}.

\section{DESCRIPTION OF THE MODEL}
\label{Desc}

We consider the Stoner-Wohlfarth particle\cite{StWo} of spherical shape characterized by the uniaxial anisotropy field $H_{a}$ and the magnetic moment $\mathbf{m} = \mathbf{m}(t)$ with a constant magnitude $|\mathbf{m}| = m$. It is assumed that the $z$ axis of the Cartesian coordinate system $xyz$ is directed along the particle easy axis and the magnetic moment $\mathbf{m}$ is under the action of both the alternating magnetic field
\begin{equation}
    \mathbf{h}(t) = h\cos(\omega t)\mathbf{e}_{x} +
    \rho h\sin(\omega t)\mathbf{e}_{y}
    \label{def_h}
\end{equation}
and the static magnetic field $\mathbf{H} = H\mathbf{e}_{z}$. Here, $\mathbf{e}_{x}$, $\mathbf{e}_{y}$, and $\mathbf{e}_{z}$ are the unit vectors along the corresponding axes of the Cartesian coordinate system, $h$ and $\omega$ are the alternating field amplitude and frequency, respectively,  and $\rho=-1,+1$ or $0$.  The case with $\rho = \pm 1$ corresponds to the circularly polarized magnetic field $\mathbf{h}(t)$ rotating in the $xy$ plane in the clockwise (if $\rho =-1$) or counterclockwise (if $\rho = +1$) direction, and $\mathbf{h}(t)$ is linearly polarized along the $x$ axis when $\rho=0$. The magnetic energy of such magnetic moment is given by
\begin{equation}
    W = -\frac{H_{a}}{2m}m_{z}^{2} - \mathbf{m}\cdot
    \mathbf{H} - \mathbf{m}\cdot\mathbf{h}(t)
    \label{def_W}
\end{equation}
($m_{\nu} = \mathbf{m} \cdot \mathbf{e}_{\nu}$, $\nu = x,y,z$) with the dot denoting the scalar product.

We first assume that the dynamics of the nanoparticle magnetic moment is governed by the stochastic LLG equation\cite{Brown, BMS}
\begin{equation}
    \frac{d\mathbf{m}}{dt} = -\gamma \mathbf{m}\times
    (\mathbf{H}_{\mathrm{eff}} +\mathbf{n}) +
    \frac{\alpha}{m}\,\mathbf{m} \times
    \frac{d\mathbf{m}}{dt}.
    \label{LLG}
\end{equation}
Here, $\gamma (>0)$ is the gyromagnetic ratio, $\alpha(>0)$ is the dimensionless damping parameter,
\begin{equation}
    \mathbf{H}_{\mathrm{eff}} = -\frac{\partial W}{\partial
    \mathbf{m}} = H_{a}\frac{m_{z}}{m}\,\mathbf{e}_{z} + \mathbf{H}
    + \mathbf{h}(t)
    \label{H_eff}
\end{equation}
is the effective magnetic field acting on the magnetic moment $\mathbf{m}$, and the cross sign denotes the vector product. As usually, the Cartesian components $n_{\nu}(t)$ of the thermal noise $\mathbf{n} = \mathbf{n}(t)$ are considered as independent Gaussian white noises characterized by zero means, $\langle n_{\nu}(t) \rangle = 0$, and correlation functions $\langle n_{\nu}(t_{1}) n_{\nu}(t_{2}) \rangle = 2 \Delta \delta(t_{2} - t_{1})$, where the noise intensity $\Delta$ is proportional to the thermal energy $k_{B}T$ ($k_{B}$ is the Boltzmann constant, $T$ is the absolute temperature), $\delta(t)$ is the Dirac $\delta$ function, and the angular brackets denote averaging over all realizations of $n_{\nu}(t)$. In general, due to the thermal fluctuations, the dynamics of $\mathbf{m}$ is stochastic, as in the case of nanoparticles in the superparamagnetic state. However, under certain conditions (see below) the nanoparticles can be single-domain and, at the same time, their magnetic moment dynamics can be approximately described by the deterministic LLG equation, i.e., Eq.~(\ref{LLG}) with $\mathbf{n}=0$. Our purpose is to determine the power loss under these conditions.

The condition that nanoparticles are in the single-domain state follows directly from the Brown's fundamental theorem.\cite{Brown2} Since, according to it, the single-domain state is energetically favorable if the nanoparticle diameter $d$ is less than a critical value $d_{\mathrm{max}}$ (which is of the order of the domain wall thickness), this condition can be written as $d < d_{\mathrm{max}}$. Next, the thermal fluctuations do not play an important role in the dynamics of $\mathbf{m}$ if the thermal energy $k_{B}T$ is much smaller than the smallest energy scale in Eq.~(\ref{def_W}). Because the condition $\widetilde{h} = h/H_{a} \ll 1$ is assumed to be realized, this energy scale is given by $mH_{a} \widetilde{h}$ or $2KV \widetilde{h}$, where $K=H_{a}M/2$ is the uniaxial anisotropy constant, $M=m/V$ is the nanoparticle magnetization, and $V=\pi d^{3}/6$ is the nanoparticle volume. From this it follows that the thermal energy can be neglected if $\kappa = 2KV \widetilde{h}/ (k_{B}T) \gg 1$. Introducing the parameter $d_{1} = [3k_{B}T/ (\pi K \widetilde{h})]^{1/3}$, interpreted as the nanoparticle diameter for which $\kappa =1$, we can rewrite the condition $\kappa \gg 1$ in the form $(d/d_{1})^{3} \gg 1$. The last inequality is satisfied with a good accuracy if $d>d_{\mathrm{min}}$, where $d_{\mathrm{min}}$ can be chosen, e.g., as $d_{\mathrm{min}} = 3d_{1}$ (in this case $\min\kappa = 27$). Thus, the nanoparticles with $d \in (d_{\mathrm{min}}, d_{\mathrm{max}})$ are single-domain and their magnetic dynamics is almost deterministic. Note that the interval $(d_{\mathrm{min}}, d_{\mathrm{max}})$ exists (i.e., $d_{\mathrm{min}} < d_{\mathrm{max}}$) if $T<T_{\mathrm{ max}}$, where $T_{\mathrm{max}}= 2KV_{ \mathrm{max}} \widetilde{h}/ (27k_{B})$ is the characteristic temperature, defined as the solution of the equation $\kappa|_{d= d_{\mathrm{max}}} = \min\kappa$ with respect to $T$, and $V_{\mathrm{max}} = \pi d_{\mathrm{max}}^{3}/6$. In other words, for each single-domain nanoparticle there is a finite temperature interval $(0, T_{\mathrm{max}})$ in which the thermal energy is negligible.

It is important to stress that even if the condition $\kappa \gg 1$ holds, there always exist the thermal fluctuations of $\mathbf{m}$ leading to a significant (of the order of $2KV \widetilde{h}$ or greater) change of the magnetic energy. But if the average time interval $\langle t \rangle$ between these fluctuations essentially exceeds the calculation time (it can be chosen as $2\pi N/ \omega$ with $N \gg 1$), they do not influence the dynamics of $\mathbf{m}$ if $\omega \gg \omega_{0}$, where $\omega_{0} = 2\pi N/ \langle t \rangle$ is the characteristic thermal frequency. Associating $\langle t \rangle$ with the mean first-passage time for the magnetic moment,\cite{DeYu} $\langle t \rangle = \sqrt{\pi/\kappa}\, e^{\kappa} (2\alpha \omega_{r})^{-1}$ ($\omega_{r}=\gamma H_{a}$ is the resonance frequency), we obtain $\omega_{0} = 4 \sqrt{\pi \kappa}\, e^{-\kappa}\alpha \omega_{r}N$. So, if $d \in (d_{\mathrm{min}}, d_{\mathrm{max}})$ and $\omega \gg \omega_{0}$ then the nanoparticles are single-domain and the dynamics of their magnetic moments can be considered as deterministic. We emphasize that these conditions are not too restrictive. In particular, according to Ref.~\onlinecite{Guim}, the Co nanoparticles at room temperature $T=300\, \mathrm{K}$ are characterized by the parameters $K = 4.12\times 10^{6}\, \mathrm{erg}/ \mathrm{cm}^{3}$, $4\pi M = 1.79\times 10^{4}\, \mathrm{G}$, and $d_{\mathrm{max}}= 96.4\, \mathrm{nm}$. Therefore, assuming that $\widetilde{h} = 0.1$, from the definition of $d_{\mathrm{min}}$ one obtains $d_{\mathrm{min}} = 13.7\, \mathrm{nm}$. Then, choosing $d=15\, \mathrm{nm}$, $\alpha = 0.1$, $N = 10^{6}$, and taking into account that in the considered case $\omega_{r} = 10^{11}\, \mathrm{s^{-1}}$, we find $\kappa = 35.2$ and $\omega_{0} = 2.2\times 10^{2}\, \mathrm{s^{-1}}$. Since $\omega_{0}$ strongly decreases with increasing $d$ (e.g., for $d=17\, \mathrm{nm}$ we have $\kappa = 51.2$ and $\omega_{0} = 2.9\times 10^{-5}\, \mathrm{s^{-1}}$), there is almost no restriction on the alternating field frequency $\omega$.

Thus, the above estimations clearly show that if the single-domain nanoparticles are not too small then the description of the dynamics of the nanoparticle magnetic moments by the deterministic LLG equation is quite justified, and this approach can be used even at room temperatures. Since we restrict ourselves to this case, below the dynamics of $\mathbf{m}$ is considered as purely deterministic.

In spherical coordinates, the deterministic LLG equation (\ref{LLG}) (when $\mathbf{n} = 0$) reduces to a system of two ordinary differential equations
\begin{eqnarray}
    (1+\alpha^{2})\dot{\theta} \!\!&=&\!\! -\, \alpha
    \sin \theta(\cos\theta + {\widetilde{H}}) + \alpha
    \widetilde{h} \cos\theta F + \widetilde{h}\, F_{\varphi},
    \nonumber\\ [6pt]
    (1+\alpha^{2})\dot{\varphi} \!\!&=&\!\! \cos\theta
    + \widetilde{H} - \widetilde{h}\cot\theta F +
    \alpha \widetilde{h} \csc\theta\, F_{\varphi},
\label{Sys1}
\end{eqnarray}
where $\theta=\theta(\tilde{t})$ and $\varphi=\varphi(\tilde{t})$ are the polar and azimuthal angles of the vector $\mathbf{m}$, respectively, $\tilde{t}=\omega_{r}t$ is the dimensionless time, the overdot denotes differentiation with respect to $\tilde{t}$, $\widetilde{H} = H/H_{a}$,
\begin{equation}
    F = \cos\varphi \cos(\widetilde{\omega} \tilde{t}) + \rho\sin
    \varphi \sin(\widetilde{\omega} \tilde{t}),
    \label{def_F}
\end{equation}
$\widetilde{\omega} = \omega/ \omega_{r}$, and $F_{\varphi} = \partial F/ \partial \varphi$.

The dynamics of $\mathbf{m}$ is accompanied by the dissipation of magnetic energy $W$. The power loss, i.e., the magnetic energy dissipation per unit time, is defined as $Q = \lim_{\tau \to \infty} (1/\tau) \int_{0}^{\tau} dtq$, where $q = -dW/dt$ is the instantaneous power loss. Since according to Eqs.~(\ref{def_W}) and (\ref{H_eff}) $q = \mathbf{H}_{\mathrm{eff}}\cdot d\mathbf{m}/dt$, the reduced power loss $\widetilde{Q} = Q/(H_{a}m\omega_{r})$, which is the quantity of our main interest, can be written in the form
\begin{equation}
    \widetilde{Q} = \lim_{\widetilde{\tau} \to \infty} \frac{1}
    {m\widetilde{\tau}} \int_{0}^{\widetilde{\tau}} d\tilde{t}\,
    \widetilde{\mathbf{H}}_{\mathrm{eff}} \cdot
    \dot{\mathbf{m}}
    \label{def_Q}
\end{equation}
with $\widetilde{\mathbf{H}}_{ \mathrm{eff}} = \mathbf{H}_{\mathrm{eff}} /H_{a}$ and $\widetilde{\tau} = \omega_{r} \tau$. To calculate $\widetilde{Q}$, we need to solve the LLG equation that, in general, can be done numerically. But in some special cases the expression for the power loss can be determined analytically.

\section{POWER LOSS: ANALYTICAL RESULTS}
\label{An_PL}

\subsection{Circularly polarized magnetic field}
\label{An_Circ}

There are two qualitatively different regimes of the steady-state dynamics of $\mathbf{m}$ in the circularly polarized magnetic field rotating about the nanoparticle easy axis, namely, periodic and quasiperiodic.\cite{BSM} The analytical results are mainly available for the periodic regime characterized by the constant precession and lag angles, $\Theta = \lim_{\tilde{t} \to \infty} \theta$ and   $\Phi = \lim_{\tilde{t} \to \infty} \phi$, where $\phi = \varphi-\rho \widetilde{\omega} \tilde{t}$. As it follows from the system of equations (\ref{Sys1}), the precession angle satisfies the equation\cite{BSM, DLHT}
\begin{equation}
    \widetilde{h}^2 = \frac{1 - \cos^{2}\Theta}{\cos^{2}
    \Theta}[(\cos\Theta + {\widetilde{H}} - \rho
    \widetilde{\omega})^{2} + (\alpha\widetilde{\omega}
    \cos\Theta)^{2}]\,
    \label{Theta}
\end{equation}
and the lag angle is connected with the precession one as follows:
\begin{equation}
    \sin\Phi = -\rho\frac{\alpha\widetilde{\omega}}
    {\widetilde{h}}\sin\Theta.
    \label{Phi}
\end{equation}
It has been shown (see also Ref.~[\onlinecite{DPL}]) that if the direction of field rotation is opposite to the direction of the natural precession of $\mathbf{m}$ then the periodic regime is always stable with respect to small perturbations. In contrast, if these directions coincide then in the parameters space there are the regions where a given periodic regime becomes unstable. Depending on the system parameters and initial conditions, the instability leads to the transition of the magnetic moment into one of three possible steady-state regimes. Two of them are periodic and one is quasiperiodic. In one of these two periodic regimes the sign of $m_{z}$ is the same as in the given periodic regime and in the other it is opposite. The transition of $\mathbf{m}$ into the periodic regime with opposite sign of $m_{z}$ corresponds to the irreversible switching of the magnetic moment (for more details see Sec.~\ref{Num_PL}).

Integrating by parts, Eq.~(\ref{def_Q}) can be represented in the form
\begin{equation}
    \widetilde{Q} =-\rho \frac{\widetilde{h} \widetilde{\omega}}
    {m} \lim_{\widetilde{\tau} \to \infty}\frac{1}{\widetilde{\tau}}
    \int_{0}^{\widetilde{\tau}} d\tilde{t}[m_{y} \cos(
    \widetilde{\omega} \tilde{t}) - \rho m_{x} \sin(
    \widetilde{\omega} \tilde{t})].
    \label{Q2}
\end{equation}
From this, writing $m_{x}$ and $m_{y}$ in spherical coordinates and using the relation $\varphi = \Phi + \rho \widetilde{\omega} \tilde{t}$ and Eq.~(\ref{Phi}), we obtain the following general expression for the reduced power loss in the case of periodic regime:
\begin{equation}
    \widetilde{Q} = \alpha \widetilde{\omega}^{2} \sin^{2}\Theta.
    \label{Q_circ}
\end{equation}
This quantity can easily be calculated at $\widetilde{h} \ll 1$. Indeed,
introducing the designation $\sigma = \mathrm{sgn}\, (\cos \Theta)$,
where $\mathrm{sgn}\,(x) $ is the sign function, and assuming that $1
+ \sigma\widetilde{H} >0$ (this assumption does not restrict the
generality of the expression below), from Eq.~(\ref{Theta}) one gets
\begin{equation}
    \Theta = \frac{\pi}{2}(1 - \sigma) + \frac{\sigma
    \widetilde{h}}{\sqrt{(1 + \sigma\widetilde{H}- \sigma\rho
    \widetilde{\omega})^{2} + (\alpha \widetilde{\omega})^{2}}}.
    \label{Theta_as}
\end{equation}
Therefore, since in the limit of small rotating field amplitudes the
condition $\sin \Theta = \widetilde{h}[(1 + \sigma\widetilde{H}- \sigma\rho
\widetilde{\omega})^{2} + (\alpha \widetilde{\omega})^{2}]^{-1/2}$ holds,
Eq.~(\ref{Q_circ}) reduces to
\begin{equation}
    \widetilde{Q} = \frac{\alpha\widetilde{h}^{2} \widetilde{\omega}^{2}}
    {(1+ \sigma\widetilde{H}- \sigma\rho \widetilde{\omega})^{2} +
    (\alpha \widetilde{\omega})^{2}}.
    \label{Q_cir}
\end{equation}

According to this result, if $\sigma\rho = -1$ or, in other words, if the direction of field rotation and the direction of the natural precession of $\mathbf{m}$ are opposite, then the reduced power loss is a monotonically increasing function of $\widetilde{\omega}$ with $\widetilde{Q} \sim \alpha\widetilde{h}^{2} \widetilde{\omega}^{2} /(1+ \sigma \widetilde{H})^{2}$ as $\widetilde{ \omega} \to 0$ and $\widetilde{Q} |_{\widetilde{ \omega}= \infty}= \alpha \widetilde{h}^{2}/(1 + \alpha^{2})$. In contrast, if $\sigma\rho = 1$, i.e., if these directions coincide, then $\widetilde{Q}$ is a unimodal function of $\widetilde{\omega}$, which at $\widetilde{\omega} = \widetilde{ \omega}_{0}$, where
\begin{equation}
    \widetilde{\omega}_{0} = 1+ \sigma \widetilde{H},
    \label{omega0_1}
\end{equation}
possesses an absolute maximum with
\begin{equation}
    \widetilde{Q} |_{\widetilde{\omega} =\widetilde{\omega}_{0}} =
    \frac{\widetilde{h}^{2}}{\alpha}.
    \label{maxQ_1}
\end{equation}
At the same time, the small-frequency behavior of $\widetilde{Q}$ and the limiting value $\widetilde{Q} |_{\widetilde{ \omega}= \infty}$ are the same as in the previous case. It should also be noted that the frequency $\widetilde{\omega}_{0}$, at which the power loss reaches the maximum, is always larger than the resonance frequency $\widetilde{ \omega}_{\mathrm{ res}} = (1+ \sigma \widetilde{H})/(1 + \alpha^{2})$, at which the precession angle (\ref{Theta_as}) becomes maximal (if $\sigma=1$) or minimal (if $\sigma=-1$).

\subsection{Linearly polarized magnetic field}
\label{An_Lin}

If the alternating magnetic field (\ref{def_h}) is linearly polarized (i.e., $\rho =0$) and its reduced amplitude $\widetilde{h}$ is small enough, then the solution of Eq.~(\ref{LLG}) with $\mathbf{n}=0$ can be represented as $\mathbf{m} = \sigma m \mathbf{e}_{z} + \mathbf{m}_{\perp}$. In the linear approximation in $\widetilde{h}$, we have $m_{\perp z} =0$,
\begin{equation}
    \widetilde{ \mathbf{H}}_{ \mathrm{eff}} = (\sigma +
    \widetilde{H}) \mathbf{e}_{z} + \widetilde{h} \cos(
    \widetilde{\omega} \tilde{t}) \mathbf{e}_{x},
    \label{Heff2}
\end{equation}
and thus the deterministic LLG equation (\ref{LLG}) reduces to
\begin{equation}
    \dot{\mathbf{m}}_{\perp} = (\sigma + \widetilde{H})
    \mathbf{e}_{z}\!\times \mathbf{m}_{\perp} + \sigma
    \alpha \mathbf{e}_{z}\! \times \dot{\mathbf{m}}_
    {\perp} - \sigma m\widetilde{h} \cos(\widetilde{\omega}
    \tilde{t}) \mathbf{e}_{y}.
    \label{LLG_lin}
\end{equation}
Since in this case $\mathbf{m}_{\perp} = m_{\perp x}\mathbf{e}_{x} + m_{\perp y} \mathbf{e}_{y}$, Eq.~(\ref{LLG_lin}) is equivalent to the system of two equations
\begin{equation}
    \begin{array}{ll}
    \dot{m}_{\perp x} = -(\sigma + \widetilde{H}) m_{\perp y}
    - \sigma\alpha \dot{m}_{\perp y},
    \\ [8pt]
    \dot{m}_{\perp y} = (\sigma + \widetilde{H}) m_{\perp x}
    + \sigma\alpha \dot{m}_{\perp x} - \sigma m\widetilde{h}
    \cos(\widetilde{\omega}\tilde{t}),
    \end{array}
\label{Sys2}
\end{equation}
whose steady-state solution can be found in the form
\begin{equation}
    \begin{array}{ll}
    m_{\perp x} = m[a\cos(\widetilde{\omega}\tilde{t}) +
    b\sin(\widetilde{\omega}\tilde{t})],
    \\ [8pt]
    m_{\perp y} = m[c\cos(\widetilde{\omega}\tilde{t}) +
    d\sin(\widetilde{\omega}\tilde{t})].
    \end{array}
\label{sol}
\end{equation}
Substituting (\ref{sol}) into (\ref{Sys2}) and using the linear independence of the trigonometric functions $\sin(\widetilde{ \omega}\tilde{t})$ and $\cos(\widetilde{ \omega} \tilde{t})$, one straightforwardly obtains
\begin{equation}
    \begin{array}{ll}
    \displaystyle a = \frac{(1 + \sigma \widetilde{H})[(1 +
    \sigma \widetilde{H})^{2} - \widetilde{\omega}^{2} + \alpha^{2}
    \widetilde{\omega}^{2}]} {[(1 + \sigma \widetilde{H})^{2} -
    \widetilde{\omega}^{2} + \alpha^{2} \widetilde{\omega}^{2}]^{2}
    + 4\alpha^{2}\widetilde{\omega}^{4}} \widetilde{h},
    \\ [12pt]
    \displaystyle b = \frac{\alpha \widetilde{\omega}[(1 +
    \sigma \widetilde{H})^{2} + \widetilde{\omega}^{2} + \alpha^{2}
    \widetilde{\omega}^{2}]} {[(1 + \sigma \widetilde{H})^{2} -
    \widetilde{\omega}^{2} + \alpha^{2} \widetilde{\omega}^{2}]^{2}
    + 4\alpha^{2}\widetilde{\omega}^{4}} \widetilde{h},
    \\ [12pt]
    \displaystyle c = -\frac{2\sigma \alpha(1 + \sigma
    \widetilde{H}) \widetilde{\omega}^{2}} {[(1 + \sigma
    \widetilde{H})^{2} - \widetilde{\omega}^{2} + \alpha^{2}
    \widetilde{\omega}^{2}]^{2} + 4\alpha^{2}\widetilde{\omega}
    ^{4}} \widetilde{h},
    \\ [12pt]
    \displaystyle d = \frac{\sigma \widetilde{\omega}[(1 +
    \sigma \widetilde{H})^{2} - \widetilde{\omega}^{2} - \alpha^{2}
    \widetilde{\omega}^{2}]} {[(1 + \sigma \widetilde{H})^{2} -
    \widetilde{\omega}^{2} + \alpha^{2} \widetilde{\omega}^{2}]^{2}
    + 4\alpha^{2}\widetilde{\omega}^{4}} \widetilde{h}.
    \end{array}
\label{abcd}
\end{equation}

According to Eq.~(\ref{Heff2}), the reduced power loss (\ref{def_Q}) in the reference case becomes
\begin{equation}
    \widetilde{Q} = \frac{\widetilde{h}}{m} \lim_{\widetilde{\tau}
    \to \infty}\frac{1}{\widetilde{\tau}} \int_{0}^{\widetilde{
    \tau}} d\tilde{t} \cos(\widetilde{\omega} \tilde{t})
    \dot{m}_{\perp x}.
    \label{Q3}
\end{equation}
From this, in accordance with Eqs.~(\ref{sol}) and (\ref{abcd}), one gets $\widetilde{Q} = \widetilde{\omega}b \widetilde{h}/2$ and
\begin{equation}
    \widetilde{Q} = \frac{\alpha\widetilde{h}^{2}}{2}\frac{\widetilde{
    \omega}^{2}[(1 + \sigma \widetilde{H})^{2} + \widetilde{
    \omega}^{2} + \alpha^{2} \widetilde{\omega}^{2}]} {[(1 +
    \sigma \widetilde{H})^{2} - \widetilde{\omega}^{2} + \alpha^{2}
    \widetilde{\omega}^{2}]^{2}+ 4\alpha^{2}\widetilde{\omega}^{4}}.
    \label{Q_lin}
\end{equation}
A simple analysis of the frequency dependence of $\widetilde{Q}$ shows that $\widetilde{Q} \sim \alpha\widetilde{h}^{2} \widetilde{\omega}^{2} /[2(1+ \sigma \widetilde{H})^{2}]$ as $\widetilde{ \omega} \to 0$, $\widetilde{Q} |_{\widetilde{ \omega}= \infty} = \alpha \widetilde{h}^{2}/[2(1 + \alpha^{2})]$, and if $\alpha \geq \sqrt{3}$ then $\widetilde{Q}$ monotonically increases with $\widetilde{\omega}$. In contrast, if $\alpha< \sqrt{3}$ then $\widetilde{Q}$ is a nonmonotonic function of $\widetilde{\omega}$, which at $\widetilde{\omega} = \widetilde{\omega}_{0}$, where
\begin{equation}
    \widetilde{\omega}_{0} = \sqrt{\frac{1}{3-\alpha^{2}}
    \bigg(1 + \frac{2}{\sqrt{1+\alpha^{2}}} \bigg)}\,
    (1+\sigma \widetilde{H}),
    \label{omega0_2}
\end{equation}
reaches the maximum value
\begin{equation}
    \widetilde{Q} |_{\widetilde{ \omega}= \widetilde{\omega}
    _{0}}\! = \frac{\alpha\widetilde{h}^{2}}{4} \frac{
    \sqrt{1+\alpha^{2}} (2 + \sqrt{1+\alpha^{2}})^{2}}
    {(\sqrt{1+\alpha^{2}} -1 + \alpha^{2})^{2} +
    \alpha^{2} (2 + \sqrt{1+\alpha^{2}})^{2}}.
    \label{maxQ_2}
\end{equation}

\begin{figure}
   \centering
   \includegraphics[totalheight=5cm] {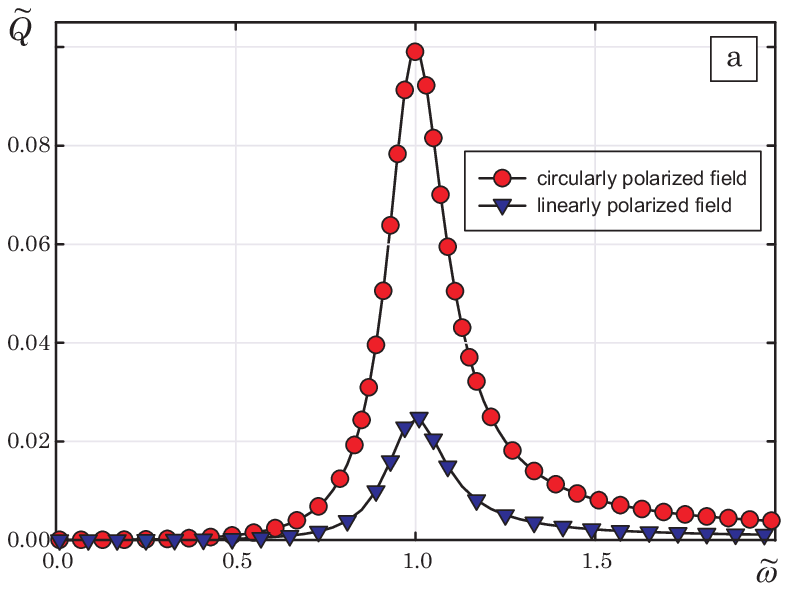}
   \includegraphics[totalheight=5cm] {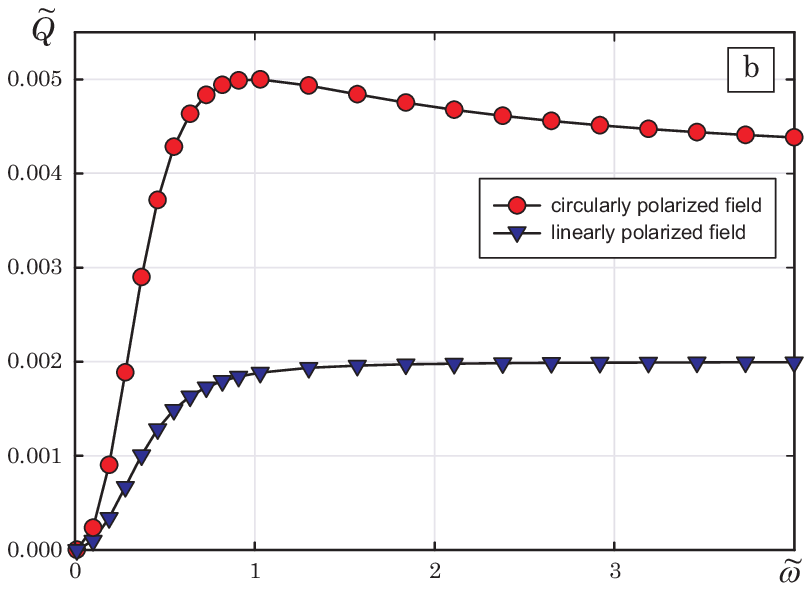}
   \caption{\label{fig:cp_lp_theor} (Color online) The reduced
   power loss $\widetilde{Q}$ as a function of the reduced frequency
   $\widetilde{\omega}$ in the small amplitude approximation. The
   parameters used in Eqs.~(\ref{Q_cir}) and (\ref{Q_lin}) are
   $\widetilde{h} = 0.1$, $\widetilde{H} = 0$, $\sigma = +1$,
   $\rho = +1$, $\alpha = 0.1$ (a) and $\alpha = 2$ (b).}
\end{figure}

Comparing the power losses in circularly and linearly polarized magnetic fields, Eqs.~(\ref{Q_cir}) and (\ref{Q_lin}), we see that in the former case $\widetilde{Q}|_{\widetilde{ \omega} \ll 1}$ and $\widetilde{Q} |_{\widetilde{ \omega}= \infty}$ are two times larger than in the latter one. Similarly, comparing Eqs.~(\ref{maxQ_1}) and (\ref{maxQ_2}), one can make sure that $\widetilde{Q} |_{\widetilde{ \omega} = \widetilde{\omega}_{0}}$ at $\alpha \ll 1$ is four times larger. The typical dependencies of $\widetilde{Q}$ on $\widetilde{\omega}$, calculated using Eqs.~(\ref{Q_cir}) and (\ref{Q_lin}), are shown in Fig.~\ref{fig:cp_lp_theor}. The case with $\alpha< \sqrt{3}$ is illustrated by Fig.~\ref{fig:cp_lp_theor}a, and the case with $\alpha> \sqrt{3}$ by Fig.~\ref{fig:cp_lp_theor}b. As seen, the power loss in nanoparticles driven by the circularly polarized magnetic field exceeds that for the linearly polarized field of the same frequency and amplitude. It can be thus concluded that in the small amplitude approximation the nanoparticle heating is more efficient in the circularly polarized magnetic field.

\section{POWER LOSS: NUMERICAL RESULTS}
\label{Num_PL}

Due to the nonlinearity of the LLG equation, with increasing the alternating field amplitude the dynamics of the magnetic moment $\mathbf{m}$ can become very complex. In particular, the regimes of quasiperiodic\cite{BSM} and chaotic\cite{APC, VaPo, BPSV} motion of $\mathbf{m}$ can be realized in circularly and linearly polarized magnetic fields, respectively. Therefore, to find the power loss in these and other cases, Eqs.~(\ref{Sys1}) should be solved numerically in a wide region of the reduced amplitudes and frequencies. To this end, one needs to perform a number of runs for each value of $\widetilde{h}$ and $\widetilde{ \omega}$ from the intervals $(\widetilde{h}_{ \mathrm{min}}, \widetilde{h}_{ \mathrm{max}})$ and $(\widetilde{\omega}_{ \mathrm{min}}, \widetilde{\omega}_{ \mathrm{max}})$ with some steps $\Delta\widetilde{h}$ and $\Delta\widetilde{\omega}$. It is important to note that the transitions between different dynamical regimes of $\mathbf{m}$ can be irreversible and can depend on the trajectory in the space of discrete variables $\widetilde{h}$ and $\widetilde{ \omega}$, as it was shown for the circularly polarized field.\cite{LPRB, DLBH} Therefore, to avoid confusion, we first fix $\widetilde{\omega}$ and then change $\widetilde{h}$ from $\widetilde{h}_{ \mathrm{min}}$ to $\widetilde{h}_{ \mathrm{max}}$ with the step $\Delta\widetilde{h}$. One run consists in finding, for given $\widetilde{h}$ and $\widetilde{ \omega}$, the solution of Eqs.~(\ref{Sys1}) on the time interval $(0, \tilde{t}_{ \mathrm{sim}})$. We use the fourth-order Runge-Kutta method and consider the simulation time $\tilde{t}_{ \mathrm{sim}}$ to be much larger than the time $\tilde{t}_{ \mathrm{loss}}$ during which the magnetic moment, exhibiting regular motion, loses memory of its initial orientation defined by the angles $\theta_{0} = \theta(0)$ and $\varphi_{0} = \varphi(0)$. In this case, we choose $\theta_{0} = 10^{-4}$, $\varphi_{0} = 0$, $\tilde{t}_{ \mathrm{sim}} = 3\cdot 10^{4}$, $\tilde{t}_{ \mathrm{loss}} = 10^{2}$, and associate the reduced power loss (\ref{def_Q}) with the numerically obtained result
\begin{eqnarray}
    \widetilde{Q} \!\!&=&\!\! \frac{1} {\tilde{t}
    _{\mathrm{sim}} - \tilde{t}_{\mathrm{loss}}} \int_{\tilde{t}_{\mathrm{loss}}}^{\tilde{t}
    _{\mathrm{sim}}} d\tilde{t}[ -(\widetilde{H}
    + \cos \theta) \sin\theta\, \dot{\theta}
    \nonumber\\ [6pt]
    &&\!\! +\, \widetilde{h}F \cos\theta\,\dot{\theta}
    + \widetilde{h}F_{\varphi} \sin\theta\,\dot{\varphi}].
\label{Q_num}
\end{eqnarray}

In the chaotic regime, the solution of Eqs.~(\ref{Sys1}) is sensitive to initial conditions. Therefore, to avoid the dependence of $\widetilde{Q}$ on these conditions arising from the finiteness of $\tilde{t}_{ \mathrm{sim}}$, an additional averaging of $\widetilde{Q}$ over $\varphi_{0}$ is performed assuming that this angle is uniformly distributed in the interval $(0, 2\pi)$.

\subsection{Small amplitude case}
\label{Num_Sm}

\begin{figure}
    \centering
    \includegraphics[totalheight=5cm] {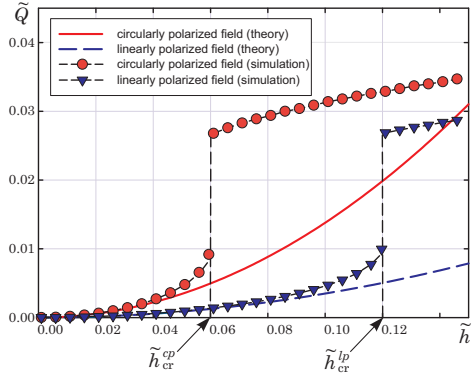}
    \caption{\label{fig:cp_lp_theor_num} (Color online)
    The reduced power loss $\widetilde{Q}$ as a function of
    the reduced amplitude $\widetilde{h}$ for circularly
    ($\rho = +1$) and linearly ($\rho = 0$) polarized
    magnetic fields. The solid and dashed lines represent
    the theoretical results obtained from Eqs.~(\ref{Q_cir})
    and (\ref{Q_lin}), respectively, and the symbols show
    the numerical results obtained from Eq.~(\ref{Q_num}).
    It is assumed that $\sigma = +1$, $\alpha = 0.1$ and
    $\widetilde{\omega} = 0.8$.}
\end{figure}

Using the numerical procedure described above, we first analyze the dependence of the power loss (\ref{Q_num}) on the alternating field amplitude for rather small values of $\widetilde{h}$. If the field frequency $\widetilde{\omega}$ is also small then $\widetilde{Q}$ as a function of $\widetilde{h}$ is well described by Eq.~(\ref{Q_cir}) (in the case of circularly polarized magnetic field) or Eq.~(\ref{Q_lin}) (in the case of linearly polarized magnetic field). But if $\widetilde{\omega}$ is relatively large then $\widetilde{Q}$ undergoes a qualitative change as $\widetilde{h}$ increases. Such behavior of the reduced power loss is demonstrated in Fig.~\ref{fig:cp_lp_theor_num}. As seen, the analytical and numerical results are almost identical for small amplitudes and, at the same time, the difference between these results becomes very pronounced even for not too large values of $\widetilde{h}$. The most remarkable feature of $\widetilde{Q}$ as a function of $\widetilde{h}$ is the existence of critical amplitudes, $\widetilde{h}_{ \mathrm{cr}}^{cp}$ and $\widetilde{h}_{ \mathrm{cr}}^{lp}$ for circularly and linearly polarized magnetic fields, respectively, at which $\widetilde{Q}$ changes abruptly. In the case of circularly polarized field, this corresponds to the so-called P-P transition from one periodic regime to another\cite{LPRB, DLBH} (see also Fig.~\ref{fig:cp_num}). The fact that $\widetilde{Q}|_{\widetilde{h} = \widetilde{h}_{ \mathrm{cr}}^{cp} - \varepsilon} < \widetilde{Q} |_{\widetilde{h} = \widetilde{h}_{ \mathrm{cr}}^{cp} + \varepsilon}$ ($\varepsilon \ll 1$) is a consequence of Eq.~(\ref{Q_circ}) and the condition $\Theta|_{\widetilde{h} = \widetilde{h}_{ \mathrm{cr} }^{cp} - \varepsilon} < \Theta|_{\widetilde{h} = \widetilde{h}_{ \mathrm{cr} }^{cp} + \varepsilon}$. The similar transition occurs also in the linearly polarized field at $\widetilde{h} = \widetilde{h}_{ \mathrm{cr}}^{lp}$. But in contrast to the previous case, the precession angle depends on time both before and after transition.

Because the behavior of $\mathbf{m}$ in circularly and linearly polarized magnetic fields with arbitrary values of $\widetilde{\omega}$ and $\widetilde{h}$ is quite different, we consider these cases separately.

\subsection{Circularly polarized magnetic field}
\label{Num_Circ}

\begin{figure}
    \centering
    \includegraphics[totalheight=5cm] {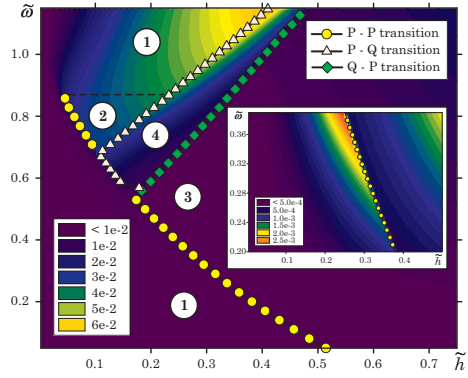}
    \caption{\label{fig:cp_num} (Color online) Color
    map for the reduced power loss $\widetilde{Q}$ and the
    diagram for the steady-state regimes of precession of
    the magnetic moment driven by the circularly polarized
    magnetic field. The regions with different periodic regimes
    of precession are indicated by numbers 1-3, and the region
    with the quasiperiodic regime of precession by number 4.
    The numerical results are obtained for $\rho = +1$,
    $\widetilde{H} = 0$ and $\alpha = 0.1$.}
\end{figure}

Using Eq.~(\ref{Q_num}), we numerically calculated the reduced power loss $\widetilde{Q}$ for a wide region in the space of parameters $\widetilde{\omega}$ and $\widetilde{h}$. The results for this quantity and the boundaries between different periodic (P) and quasiperiodic (Q) regimes of the steady-state precession of the magnetic moment are shown in Fig.~\ref{fig:cp_num} for $\widetilde{H} = 0$. Region 1 represents the periodic regime of precession of $\mathbf{m}$, which is described by Eqs.~(\ref{Theta}) and (\ref{Phi}) with $\Theta < \pi/2$ (i.e., $\sigma = +1$) and $\rho = +1$. When $\widetilde{h}$ increases, this precession becomes unstable\cite{DLHT} and, depending on the reduced frequency, the magnetic moment can make a transition to one of three steady states. These states are respectively characterized by (i) periodic precession with $\Theta < \pi/2$ (region 2), (ii) periodic precession with $\Theta > \pi/2$ (region 3), and (iii) quasiperiodic precession (region 4). The precession angle $\Theta$ as a function of $\widetilde{h}$ is discontinuous at the boundaries (denoted by circles) between regions 1 and 2 ($\Theta|_{1} < \Theta|_{2} < \pi/2$) and between regions 1 and 3 ($\Theta|_{1} < \pi/2 < \Theta|_{3}$). An important difference between these P-P transitions is that the former is reversible, while the latter, corresponding to the magnetic moment switching, is irreversible.

\begin{figure}
    \centering
    \includegraphics[totalheight=3.5cm] {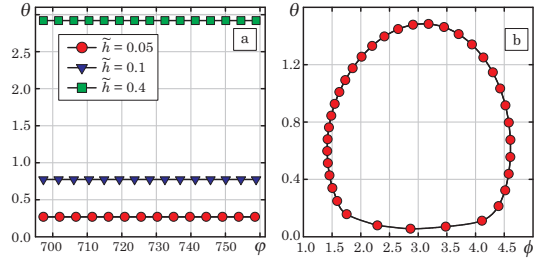}
    \caption {\label{fig:cp_traj} (Color online)
    Examples of steady-state trajectories of the
    magnetic moment driven by the circularly polarized
    magnetic field. The simulation parameters are
    chosen to be $\rho = +1$, $\widetilde{H} = 0$,
    $\alpha = 0.1$ and $\widetilde{\omega}=0.8$.
    Trajectories for periodic regimes in regions 1
    ($\widetilde{h} = 0.05$), 2 ($\widetilde{h} =
    0.1$) and 3 ($\widetilde{h} = 0.4$), and for the
    quasiperiodic regime in region 4 ($\widetilde{h}
    = 0.26$) are shown in (a) and (b), respectively.}
\end{figure}

In region 4, the steady-state precession of $\mathbf{m}$ is quasi\-periodic, i.e., the angles $\theta$ and $\phi = \varphi-\rho \widetilde{\omega} \tilde{t}$ are periodic functions with the same period, which in general is not commensurable with the field period $2\pi/ \widetilde{\omega}$. The transition from the periodic regime of precession in region 1 or 2 to the quasiperiodic regime in region 4 (P-Q transition, triangle line) is reversible. In contrast, the transition from the quasiperiodic regime of precession to the periodic regime in region 3 (Q-P transition, square line) is irreversible. As an illustration, in Fig.~\ref{fig:cp_traj} we show the steady-state trajectories of $\mathbf{m}$ in regions 1-3 (a) and in region 4 (b).

The numerical results for the reduced power loss $\widetilde{Q}$ are show in Fig.~\ref{fig:cp_num} as a color map. For the periodic regimes (regions 1-3), these results are in excellent agreement with those obtained from Eq.~(\ref{Q_circ}). We note the following features of the reduced power loss. First, since $\Theta|_{1} < \Theta|_{2}$, the P-P transition to region 2 is accompanied by an abrupt increase of $\widetilde{Q}$. Second, the transition to region 3 is followed by an abrupt decrease of $\widetilde{Q}$ (see also inset in Fig.~\ref{fig:cp_num}). Because after transition to region 3 the direction of the natural precession of $\mathbf{m}$ becomes opposite to the direction of field rotation, this occurs for both the P-P and Q-P transitions. Third, the P-Q transition from region 1 or 2 to region 4 does not lead to a discontinuity in $\widetilde{Q}$. And fourth, as is clearly seen from this figure, the maximum of the reduced power loss is reached near the (triangle) line of the P-Q transition.

\begin{figure}
    \centering
    \includegraphics[totalheight=5cm] {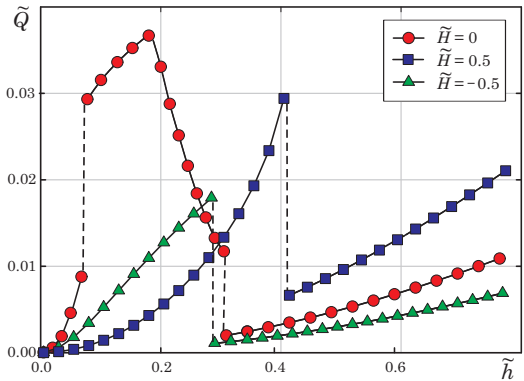}
    \caption {\label{fig:cp_Q_with_Hz} (Color online)
    The reduced power loss $\widetilde{Q}$ as a function
    of the reduced amplitude $\widetilde{h}$ of circularly
    polarized magnetic field for different values of the
    reduced magnetic field $\widetilde{H}$. The parameters
    $\rho$, $\alpha$ and $\widetilde{\omega}$ are the same
    as in Fig.~\ref{fig:cp_traj}.}
\end{figure}

It should also be noted that the static magnetic field changes the magnetic moment dynamics and hence influences the power loss. For illustration, in Fig.~\ref{fig:cp_Q_with_Hz} we show the dependence of $\widetilde{Q}$ on $\widetilde{h}$ for different values of $\widetilde{H}$. The jumps of $\widetilde{Q}$ at $\widetilde{H} =0$ correspond to the P-P (reversible) and Q-P (irreversible) transitions, while the jumps at $\widetilde{H} =\pm 0.5$ correspond to the P-P (irreversible) transitions associated with switching of $\mathbf{m}$.

\subsection{Linearly polarized magnetic field}
\label{Num_Lin}

\begin{figure}
    \centering
    \includegraphics[totalheight=3.5cm] {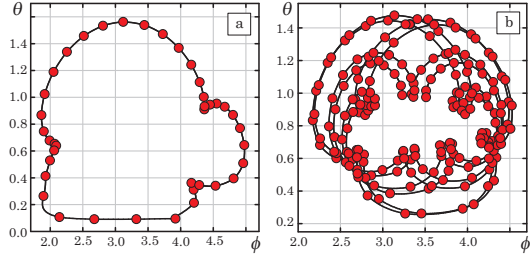}
    \caption {\label{fig:lp_traj} (Color online)
    Examples of regular (a) and chaotic (b) trajectories
    of the magnetic moment driven by the linearly polarized
    magnetic field. The simulation parameters are $\rho = 0$,
    $\widetilde{H} = 0$, $\alpha = 0.1$, $\widetilde{\omega}
    =0.8$, $\widetilde{h} = 0.4$ (a) and $\widetilde{h} =
    0.46$ (b).}
\end{figure}

In this field, the dynamics of the magnetic moment differs considerably from that described in the previous section. One of the differences is the absence of the  periodic regime of precession of $\mathbf{m}$ in the above sense. But the most striking difference is that the linearly polarized magnetic field can induce the chaotic regime of precession.\cite{APC, VaPo, BPSV} This implies the existence of regions in the parameter space in which the regular dynamics of $\mathbf{m}$ is still very complex (pre-chaotic behavior). For comparison, in Fig.~\ref{fig:lp_traj} we plot the trajectories of $\mathbf{m}$ performing regular and chaotic (in a finite time interval) precessions.

Using the previously described procedure, we can compute the reduced power loss $\widetilde{Q}$ in the case of linearly polarized magnetic field as well. To analyze its dependence on the character of precession, we need to establish  whether the precession is regular or chaotic for a given set of parameters. This can be done by determining the sign of the largest Lyapunov exponent $\lambda_{1}$ ($\lambda_{1}>0$ corresponds to the chaotic behavior), which describes the divergence of neighboring trajectories.\cite{WSSV, RaSr} For our system, this quantity can be introduced as
\begin{equation}
    \lambda_{1}=\lim_{k \to \infty}\frac{1}
    {k\, \Delta\widetilde{\tau}}
    \sum_{n=1}^{k}\ln\sqrt{[\delta\theta(n\,
    \Delta\widetilde{\tau})]^{2} +[\delta
    \varphi(n\, \Delta\widetilde{\tau})]^{2}},
    \label{Lyap_exp}
\end{equation}
where $\Delta\widetilde{\tau}$  is a short time interval $(\Delta\tilde{t} \ll \Delta\widetilde{\tau} \ll \tilde{t}_{\mathrm{sim}})$, $\Delta\tilde{t}$ is the simulation time step, and $\delta \theta$ and $\delta \varphi$ satisfy the  system of linear equations
\begin{equation}
    \begin{array}{ll}
    (1+\alpha^{2})\delta \dot{\theta} = f_{\theta}
    \delta\theta + f_{\varphi} \delta\varphi,
    \\ [8pt]
    (1+\alpha^{2})\delta\dot{\varphi} = g_{\theta}
    \delta\theta + g_{\varphi} \delta\varphi.
    \end{array}
\label{Sys_lin}
\end{equation}
Here, $f$ and $g$ are the right hand sides (at $\rho=0$) of the first and second equations in (\ref{Sys1}), respectively, and the indexes $\theta$ and $\varphi$ denote differentiation with respect to these variables. To calculate $\lambda_{1}$, we first solve Eqs.~(\ref{Sys1}) on the interval $(0, \tilde{t} _{\mathrm{sim}})$. Then, using the same initial conditions for $\delta\theta$ and $\delta\varphi$ at $\tilde{t} = \tilde{t}_{n}$ ($\tilde{t}_{n} = (n-1)\Delta \widetilde{ \tau},\, n= \overline{1,k}$), we solve Eqs.~(\ref{Sys_lin}) on the intervals $(\tilde{t}_{n}, \tilde{t}_{n} + \Delta\widetilde{\tau})$ and from Eq.~(\ref{Lyap_exp}) find $\lambda_{1}$.

\begin{figure}
    \centering
    \includegraphics[totalheight=5cm] {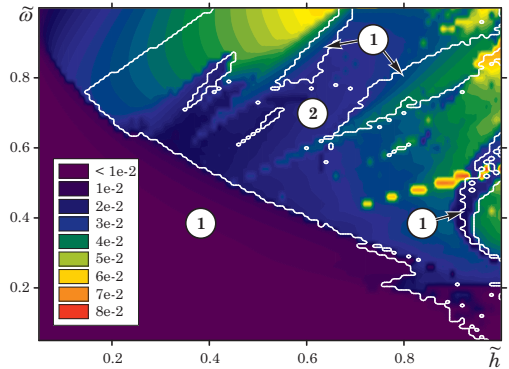}
    \caption {\label{fig:lp_num} (Color online) Color map
    for the reduced power loss $\widetilde{Q}$ and the regions
    of regular (1) and chaotic (2) precession of the magnetic
    moment driven by the linearly polarized magnetic field.
    The change of sign of the largest Lyapunov exponent is
    indicated by white lines. The parameters $\rho$,
    $\widetilde{H}$ and $\alpha$ are the same as in
    Fig.~\ref{fig:lp_traj}.}
\end{figure}

The reduced energy loss and the lines on which the largest Lyapunov exponent $\lambda_{1}$ changes sign, obtained for $\Delta \tilde{t} = 2\cdot 10^{-5}$, $\Delta\widetilde{\tau} = 2\cdot 10^{-3}$, $k = 1.5\cdot 10^{7}$, $\delta \theta|_{\tilde{t}_{n}} = 1$ and $\delta \varphi|_{\tilde{t}_{n}} = 0$ (this is the usual choice of the initial conditions for $\delta \theta$ and $\delta \varphi$), are shown in Fig.~\ref{fig:lp_num}. We note three important features of these results. First, the regions in the $\widetilde{h}$-$\widetilde{ \omega}$ plane with regular $(\lambda_{1} <0)$ and chaotic $(\lambda_{1} >0)$ dynamics of $\mathbf{m}$ are distributed very unevenly. We remark in this context that if a contour connecting two points on this plane crosses the white lines an even (odd) number of times, then the character of the magnetic moment dynamics in these points is the same (different). It should be emphasized, however, that the results concerning the regions with regular and chaotic dynamics of the magnetic moment should be considered as preliminary. The reason is that, due to the existence of two equilibrium directions of the magnetic moment in uniaxial nanoparticles and rather large $\delta \theta|_{ \tilde{t}_{n}}$, the condition $\lambda_{1}>0$ may be expected to hold for some regular trajectories as well. In other words, the condition $\lambda_{1}> 0$ may appear as an artifact of the numerical scheme. The analysis of the long-time behavior of $\mathbf{m}$  confirms the existence of such trajectories in some regions of the $\widetilde{h}$-$\widetilde{ \omega}$ plane (these regions are not shown in Fig.~\ref{fig:lp_num}). Second, the reduced power loss can experience an abrupt change not only under transitions between regular and chaotic regimes of precession, but also under transitions between different regimes of regular precession (see also Fig.~\ref{fig:cp_lp_theor_num}) and between different regimes of chaotic precession. And third, comparing in Figs.~\ref{fig:cp_num} and \ref{fig:lp_num} the values of the reduced power loss, one can conclude that nanoparticle heating in the circularly polarized magnetic field, whose amplitude and frequency are close to the line of P-Q transitions, is more efficient than in the linearly polarized field. On the other hand, if the reduced amplitude $\widetilde{h}$ is large enough, then the linear polarization of the alternating magnetic field is more preferable for heating purpose.

\section{SUMMARY AND CONCLUSIONS}
\label{Summ}

Using the deterministic Landau-Lifshitz-Gilbert equation, we have studied in detail the dependence of the reduced power loss and the character of forced precession of the nanoparticle magnetic moment on amplitude and frequency of circularly and linearly polarized magnetic fields. The circularly polarized field, whose plane of polarization is perpendicular to the anisotropy axis, can generate three periodic regimes of forced precession (two of them occur in the up state of the magnetic moment, and the other occurs in the down state) and one quasiperiodic regime. We have determined the regions in the amplitude-frequency plane where these regimes exist and have calculated the power loss inside them. A remarkable feature of the power loss is that it changes abruptly at some boundaries between these regions. In particular, the transition from the regions with periodic or quasiperiodic precession to the region with periodic precession in the down state is accompanied by an abrupt decrease of the power loss. In contrast, the transition between regions with different periodic precessions in the up state is accompanied by an abrupt increase of the power loss if the precession angle increases under transition. We have also established that the power loss reaches the largest values near the boundary between the region with periodic precession of the magnetic moment in the up state and the region with quasiperiodic precession. This is the condition under which the nanoparticle heating by the circularly polarized magnetic field is the most efficient.

The linearly polarized field, whose axis of polarization is perpendicular to the anisotropy axis, can induce both regular and chaotic regimes of precession of the magnetic moment. By analyzing the largest Lyapunov exponent and the long-time behavior of the magnetic moment, we have delimited the regions in the amplitude-frequency plane where the magnetic moment exhibits the regular and chaotic behavior. The distribution of these regions has a complex character and the power loss corresponds, in general, this distribution. Nevertheless, the transitions between different regimes of regular and chaotic precession can also strongly affect the power loss. Thus, our results provide evidence that the energy dissipation in single-domain nanoparticles crucially depends on the character of the magnetic moment dynamics.

\section*{ACKNOWLEDGMENTS}

T.V.L. and A.Yu.P. acknowledge the support of the Cabinet of Ministers of Ukraine, The Program of Studying and Training for Students, PhD Students, and Professor's Stuff Abroad (order dated April 13, 2011 No 411). T.V.L. and S.I.D. appreciate the support of the Ministry of Education and Science of Ukraine (Project No 0112U001383). The authors are also grateful to Yu.~S.~Bystrik for the assistance in the determination of the largest Lyapunov exponent.

\end{document}